\documentclass[preprint,showpacs,preprintnumbers,amsmath,amssymb]{revtex4}

\usepackage{dcolumn}
\usepackage{bm}
\usepackage{graphicx,subfigure}

\begin{document}

\title{Synthetic Diamond and Wurtzite Structures Self-Assemble with
Isotropic Pair Interactions}

\author{Mikael C. Rechtsman$^1$} \author{Frank H. Stillinger$^2$}
\author{Salvatore Torquato$^{2,3,4}$} \affiliation{$^1$Department of
Physics, Princeton University, Princeton, New Jersey, 08544}
\affiliation{$^2$Department of Chemistry, Princeton University,
Princeton, New Jersey, 08544} \affiliation{$^3$Program in Applied and
Computational Mathematics and PRISM, Princeton, New Jersey, 08544}
\affiliation{$^4$Princeton Center for Theoretical Physics, Princeton,
New Jersey, 08544}
\date{\today}

\begin{abstract}
Using inverse statistical-mechanical optimization techniques, we have
discovered isotropic pair interaction potentials with strongly
repulsive cores that cause the tetrahedrally-coordinated diamond and
wurtzite lattices to stabilize, as evidenced by lattice sums, phonon
spectra, positive-energy defects, and self-assembly in classical
molecular dynamics simulations.  These results challenge conventional
thinking that such open lattices can only be created via directional
covalent interactions observed in Nature.  Thus, our discovery adds to
fundamental understanding of the nature of the solid state by showing
that isotropic interactions enable the self-assembly of open crystal
structures with a broader range of coordination number than previously
thought.  Our work is important technologically because of its direct
relevance generally to the science of self-assembly and specifically
to photonic crystal fabrication.

\end{abstract}

\maketitle

\section{Introduction}

The science of ``self-assembly", as introduced by George Whitesides
and coworkers about fifteen years ago \cite{whitesides2} as a
promising research direction, is a large and rapidly growing field of
tremendous technological potential and fundamental interest.  Its
motivating factor is simple: direct micro- and nano-fabrication of
circuitry (electronic, optoelectronic, or others) becomes
prohibitively time-consuming and/or expensive below a certain length
scale, and as a result, scientists and engineers are searching for
particles on the mesoscopic scale that by themselves assemble into
potentially useful structures by virtue of their mutual interactions;
hence the term `self-assembly'.  There is a large number of examples
of theoretical, experimental, and computational studies on this topic
\cite{Jen99,chaikin1,Stellacci,chaikin2,geissler1}.  Colloidal
systems command particular interest, as interaction potentials in these
systems have become increasingly tailorable with the advent of new
ways to functionalize the colloidal surface.  These systems have
tremendous capacity to assemble exotic lattices and hence to yield
photonic bandgap structures.  In particular, the diamond lattice is
known to have a pronounced photonic bandgap \cite{photonics}.

In this paper, we obtain the fundamentally and technologically
important result that the three-dimensional diamond and wurtzite
structures can self-assemble with isotropic interactions possessing a
stronly repulsive core, and present the corresponding potential
functions, which have been derived using optimization techniques
developed previously \cite{RST-PRL1}.  Specifically, we show that an
$N$-body classical system with particles interacting via one of our
derived isotropic potentials has as its ground state the corresponding
lattice, in a specific volume (or density) range.  Unlike previous
attempts to solve this problem, we come to this conclusion only after
satisfying several important criteria: (1) lattice sums show that
there is a positive pressure range in which the given lattice is
stable; (2) all crystal normal mode frequencies are real at specific
volume $v$; (3) defects (vacancies and interstitials) are shown to
cost the system energy; and (4) the system self-assembles in a
molecular dynamics simulation that starts above the freezing point and
is slowly cooled.  In (4), the system may start from an entirely
random configuration or with a layer of fixed particles to promote
epitaxial growth.  Hence we make the important distinction here
between homogeneous and heterogenous nucleation in self-assembly.  It
is of course a more stringent requirement that the desired lattice
self-assemble from a random configuration (homogeneous).  We find that
the diamond crystal assembles from a completely random configuration,
but the wurtzite crystal requires an epitaxial layer for it to do so (heterogeneous).

The importance of criterion (4) to establish a ground state
structure has been generally overlooked in previous work.  Currently,
most studies rely exclusively on criteria (1) and/or (2) to determine
lattice stability.  The only result that is guaranteed by satisfying
(1) and (2) is that the structure is a local minimum, not a global
one.  Computationally, the only practical way that one can ensure that
the desired target structure is the ground state is by doing the
cooling and annealing step, (4).  However, if the presumed ground
state is not achieved by an annealing, it does not necessarily imply
that it is not the ground state if the resulting energy of the
structure produced exceeds that of the target structure.  There may
simply be a kinetic bottleneck that prevents the annealing process
from finding the ground state in a reasonable amount of computer
time. 

The approach by theorists and experimentalists towards finding complex
self-assembling materials has primarily been {\it Edisonian} (i.e.,
trial-and-error) in nature rather than being systematic and deductive.
Motivated by the increasing control that experimentalists have over
interparticle interactions, in colloids in particular, we have
previously derived a systematic procedure to optimize these
interactions to produce desired structures.  This is an example of
``inverse statistical mechanics'': Instead of using the interactions
in a system to find the resulting behavior, we target a particular
behavior (e.g. structure), and derive optimal interactions to yield
such a result.  We have recently applied these schemes to
two-dimensional monodisperse, isotropic systems \cite{RST-PRL1}, and
in the present paper, we use one of these methods, the so-called
``zero-temperature scheme'', to derive isotropic interaction
potentials that permit the self-assembly of the diamond and wurtzite
lattices respectively. The self-assembly of any low-coordinated target
structure in three dimensions is considerably more challenging than in
two dimensions because the number of possible competing structures
grows dramatically with increasing dimension.

With the exception of the so-called patchy particles \cite{Glotzer},
colloidal interactions are isotropic, and thus we believe it is a
technologically relevant as well as theoretically intriguing question
to ask: What are the limits of isotropic interaction potentials for
self-assembly?  Our discovery adds to fundamental understanding of
the nature of the solid state by showing that isotropic interactions
enable the self-assembly of crystal structures with a much broader
range of coordination numbers than previously thought.

The self-assembly of a diamond lattice of dielectric spheres has been
something of a holy grail in colloid science, since it was found that
in the diamond lattice, unlike in the face-centered cubic (fcc) and
body-centered cubic (bcc) lattices, for example, there exists a
bandgap in the photonic spectrum \cite{photonics}.  This is the
central necessary condition for and enabler of photonic circuitry,
since this property allows control over passage vs. blockage of light,
in analogy to how transistors control electric current
\cite{diamondselfassembly}.  It should be noted that there has been
recent work on icosahedral quasicrystals as desirable photonic bandgap
structures \cite{ChaikinSteinhardt}.  While this is a promising avenue
of research, its central stumbling block is that self-assembly of
quasicrystals in three dimensions is extremely difficult.  Indeed, a
single defect destroys strict quasicrystallinity.  Furthermore, there
is an ongoing controversy over whether a quasicrystal can be assembled
using only local information.  The diamond-structure potential we find
is radially complex, and seemingly difficult to fabricate in the lab
directly.  That said, technology that allows experimentalists to
control colloidal and nanoparticle interaction has been advancing
significantly in recent years, with DNA functionalization
\cite{chaikin1}, for example, becoming a very useful method.
Furthermore, radial complexity in the interaction potential is found
in potentials of mean force \cite{potentialsofmeanforce}.  Recently,
experimentalists have observed diamond-like behavior in mixed silver
and gold nanoparticle systems \cite{grzybowski1}, however the precise
form of the interparticle interaction potential is currently unknown.

Watzlawek, Likos, and L\"owen (WLS) have devised an isotropic
interaction potential that models star polymer systems, with good
agreement with experiment \cite{Likos}.  In particular, the WLS group
has found, for certain of their parameters, a region of stability for
the diamond lattice, and thus an isotropic potential that favors it.
The type of potential used in the WLS study has a soft core; in
particular, for small $r$, $V(r)\sim-\log(r)$.  While their potential
is structurally simpler than the one we present here, it is not a
potential that can be considered analogous to those in colloidal
systems with present technology because colloidal particles have too
hard a core.  We have discovered, in the course of performing
optimization runs, that assembling open lattices with hard core
potentials requires the potential to have more features than that of
the WLS potential.

We believe that tetrahedrally coordinated self-assembly via an
isotropic potential is a counterintuitive result because conventional
wisdom, based on atomic and molecular chemistry, has biased thinking
towards the belief that highly anisotropic interactions are necessary
for such locally directional structure as in diamond or wurtzite, as
in the case of the carbon-carbon covalent bonds in the true diamond
crystal.  The present work constitutes a clear demonstration that this
chemical intuition is too restrictive.  We have previously found an
isotropic potential that causes spontaneous assembly of the 6-fold
coordinated simple cubic lattice \cite{RST-PRE2}, but the tetrahedral
lattices present a significantly greater challenge.  Finding such
potentials is difficult, as the diamond and wurtzite lattices are
close structural competitors with one another (and therefore close in
energy for short-ranged potentials).

In order to understand how the diamond lattice is energetically
distinguished from other lattices, consider wurtzite as a competitor
to diamond in a system with an isotropic interaction potential.
Whereas the diamond lattice can be considered as two interlaced fcc
lattices, the wurtzite is two interlaced hcp lattices.  Both are
tetrahedrally coordinated.  For the sake of argument, call the nearest
neighbor distance $a$ in both lattices.  Both have 12-fold coordinated
neighbors at $\sqrt{8/3}a\sim 1.63a$, but the wurtzite has a single
extra neighbor at $(5/3)a$.  Both have neighbors at $\sqrt{11/3}a$,
but the diamond is 12-fold coordinated there where the wurtzite is
9-fold.  It is evident that energetically differentiating between
these two lattices (and causing one to self-assemble in a simulation
rather than the other) using an isotropic, short ranged potential is
challenging.  Table 1 is gives the diamond and wurtzite coordination
numbers for their first few coordination shells.\\

\begin{table}
\centering
\begin{tabular}{l|l|l}
$r/a$&diamond&wurtzite\\
\hline
\hline
1&4&4\\
$\sqrt{8/3}$&12&12\\
5/3&0&1\\
$\sqrt{11/3}$&12&9\\
$\sqrt{16/3}$&6&6\\
7/3&0&6\\
$\sqrt{19/3}$&12&9\\
8/3&0&2 \\
$\sqrt{8}$&24&18\\

\end{tabular}
 \caption{Coordination structure of the diamond and wurtzite lattices.
 Here, we set $a$ to be the distance to the nearest neighbor.  If
 both lattices have the same specific volume, $a$ for diamond is
 equal to $a$ for wurtzite.}
\end{table}

In Section II, we briefly discuss the optimization methodology,
and then proceed in Section III to a discussion of the results for the
diamond and wurtzite cases.  In this section, we report on all
necessary criteria: lattice sums, phonons, defect energies, and MD
self-assembly. We also include a discussion of the melting behavior of
each crystal.  We conclude in section IV with some final comments
about this approach.

\section{Methodology}

The inverse method used to obtain the interaction potentials, $V_D(r)$
and $V_W(r)$, is discussed extensively in our previous work
\cite{RST-PRE1}, and so will not be discussed here at any length.  It
is, however, important to note that this ``zero-temperature scheme''
optimizes the energetic stability of a given target lattice (in this
case, diamond and wurtzite), in comparison to competitor lattices (chosen
previously), subject to the condition that the target lattice is linearly
mechanically stable at an appropriately chosen specific volume
(inverse density).  Linear stability is tested while the optimization
is being carried out by calculating the phonon spectrum, and thus
making sure that every normal mode of the crystal, at any given wave
vector in the Brillouin zone (BZ) is real.  The competitor lattices
used were the fcc, bcc, hexagonal close-packed (hcp), simple hexagonal
(shex), and simple cubic (sc) structures.

Of course the optimization scheme for the isotropic interaction cannot
search over the entire space of functions of a single variable,
therefore we choose a suitably normalized, well-motivated family of
potentials defined by a finite number of parameters,
$\{a_0,...,a_n\}$, and find the set of $a_i$'s that maximizes the
energetic difference between the target and competitor lattices
subject to the constraint of linear mechanical stability at a
particular specific volume.

\section{Results}

\noindent {\it A. Diamond}\\

In this case, a parameterization was chosen such that the potential
has three local minima, with ratios of the positions of the second to
the first, and of the third to the first, being $\sqrt{8/3}$ and
$\sqrt{11/3}$ respectively.  These are the exact distance ratios of
the neighbors in the diamond lattice.  The target coordination numbers
for the first three neighbors are 4, 12, and 12, and thus the first
local minimum was constrained to be positive in the parameterization,
in order to discriminate against closed packed and the body-centered
cubic lattices, which have coordination numbers 12 and 8
respectively.  It should be noted that a number of different
parameterizations were attempted, many with simpler functional forms
(i.e. with fewer local minima and maxima), but the potential we report
here is by all criteria the most suitable.  It is

\begin{equation}
V_D(r) = \frac{5}{r^{12}}-\frac{6}{r^{10}} + 17.0\exp\left[-2.5r\right]\\
-0.362477\exp\left[-150(r-1.702)^2\right] -
0.263218\exp\left[-166(r-1.998)^2\right].
\end{equation}         

\begin{figure}
\includegraphics[width=10cm]{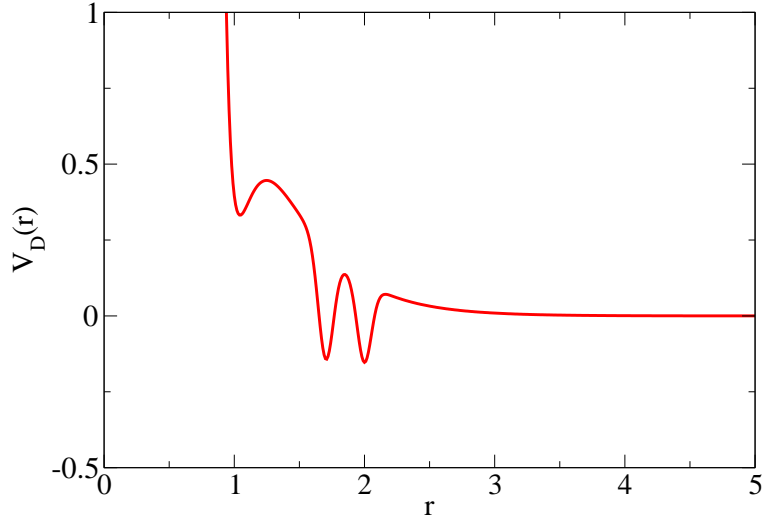}
\caption{(Color online) Functional form of the potential, $V_D(r)$, given in Eq. 1.  Given in reduced units.}
\label{fig:potential}
\end{figure}

This potential is shown in Fig. \ref{fig:potential}.  It consists of a
Lennard-Jones-type term (a 12-10 Lennard-Jones), an exponential decay,
and two Gaussians that serve to localize the second and third
neighbors for the sake of mechanical stability of the lattice.  It is
important to note here that this potential defines the energy and
length units that we use for this study.  Thus, derived quantities,
such as pressure and volume, have units determined by the length and
energy units of this potential.

In the optimization scheme, the specific volume is chosen for us: it
is that which places the nearest neighbor of the diamond lattice
exactly at the first local minimum.  This occurs at $r=1.0450$, which
sets the specific volume to be $1.75694$.  

\begin{figure}
\includegraphics[width=10cm]{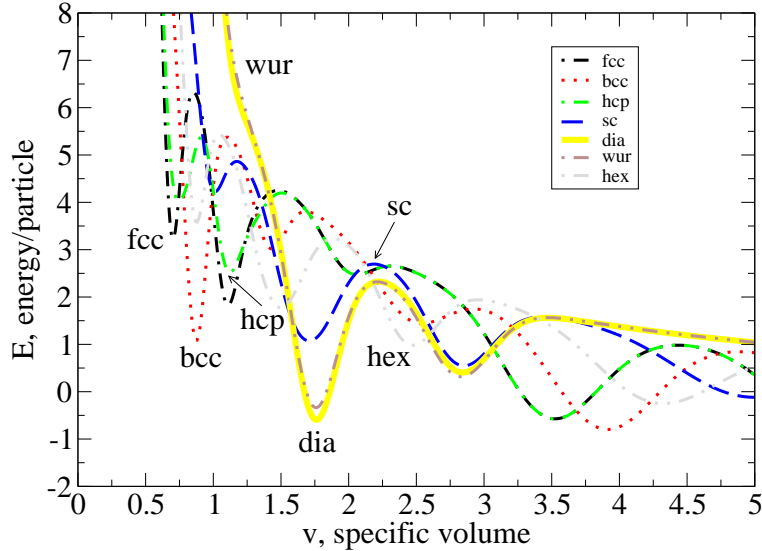}
\caption{(Color online) Lattice sums of seven lattices, including diamond, using the
potential given in Eq. 1.  The double-tangent construction determines that
at zero temperature, the pressure range of stability for the diamond
lattice is 0.12 through 1.89.  Here, ``dia'' refers to diamond, ``wur'' to wurtzite, and ``hex'' to simple hexagonal (with $c/a$=1).  Given in reduced units.  }
\label{fig:latticesums}
\end{figure}

An essential ingredient of the optimization scheme is the lattice
sums, which give the Madelung energies (crystal energies) of a number
of lattices, including diamond, over a range of specific volume.
These are shown in Fig. \ref{fig:latticesums}.  The lattice sums not
only show that the diamond lattice has the lowest energy of all of its
competitors at the chosen $v$, but also the double-tangent
construction yields the zero-temperature range of pressures in which
the diamond lattice is stable.  That range is 0.12 through 1.89, in
terms of the units defined by the potential (Eq. 1).

\begin{figure}

\subfigure[View 1: Shows triangular lattice slice of diamond.]  
{
\includegraphics[width=5cm]{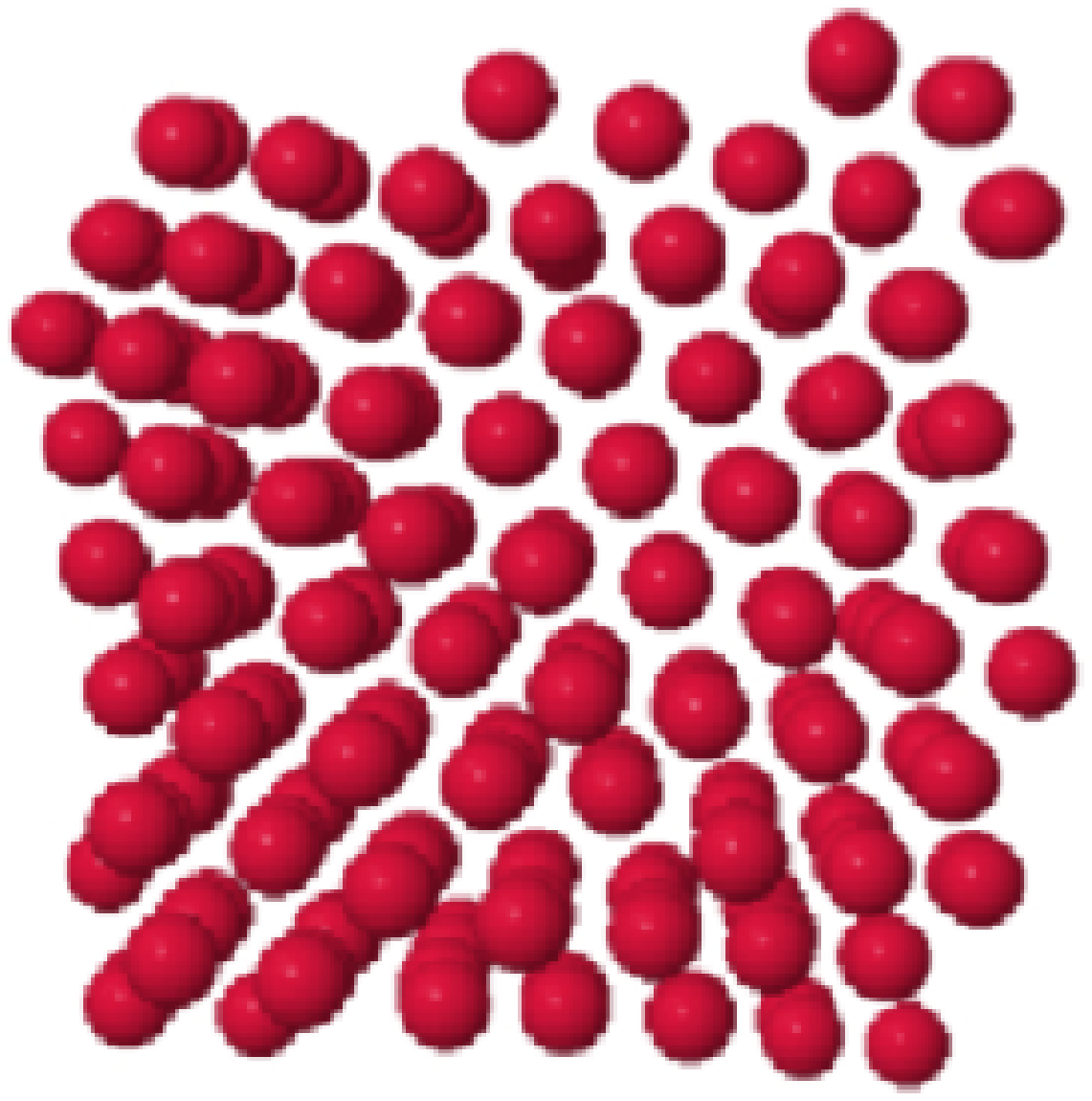}
}
\subfigure[View 2: ``Hexagons'' in this view are indicative of the
diamond structure.]  
{
\includegraphics[width=5cm]{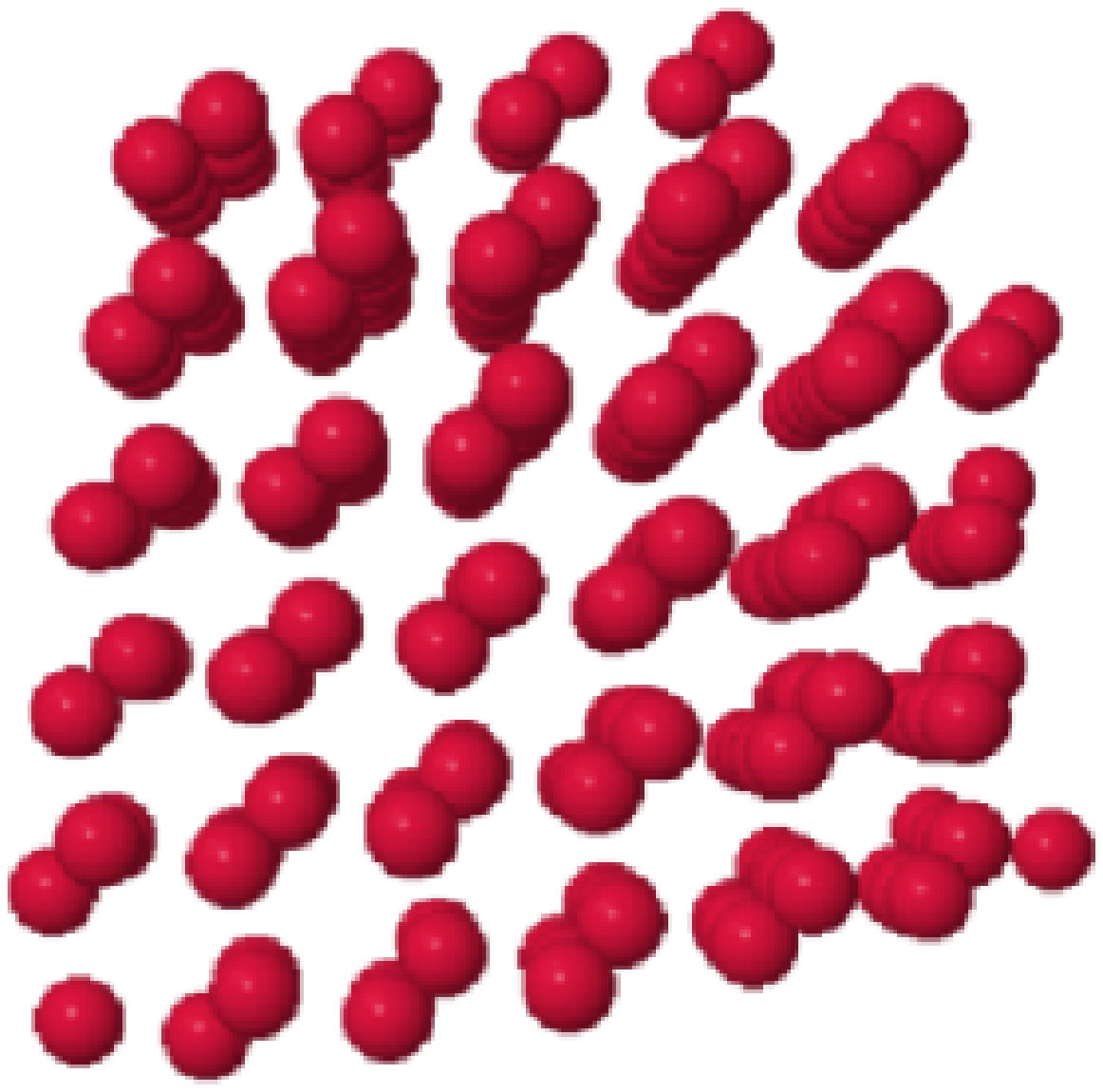}
}
\subfigure[View 3: Square lattice shows most
clearly the cubic symmetry of the diamond lattice.] 
{
\includegraphics[width=5cm]{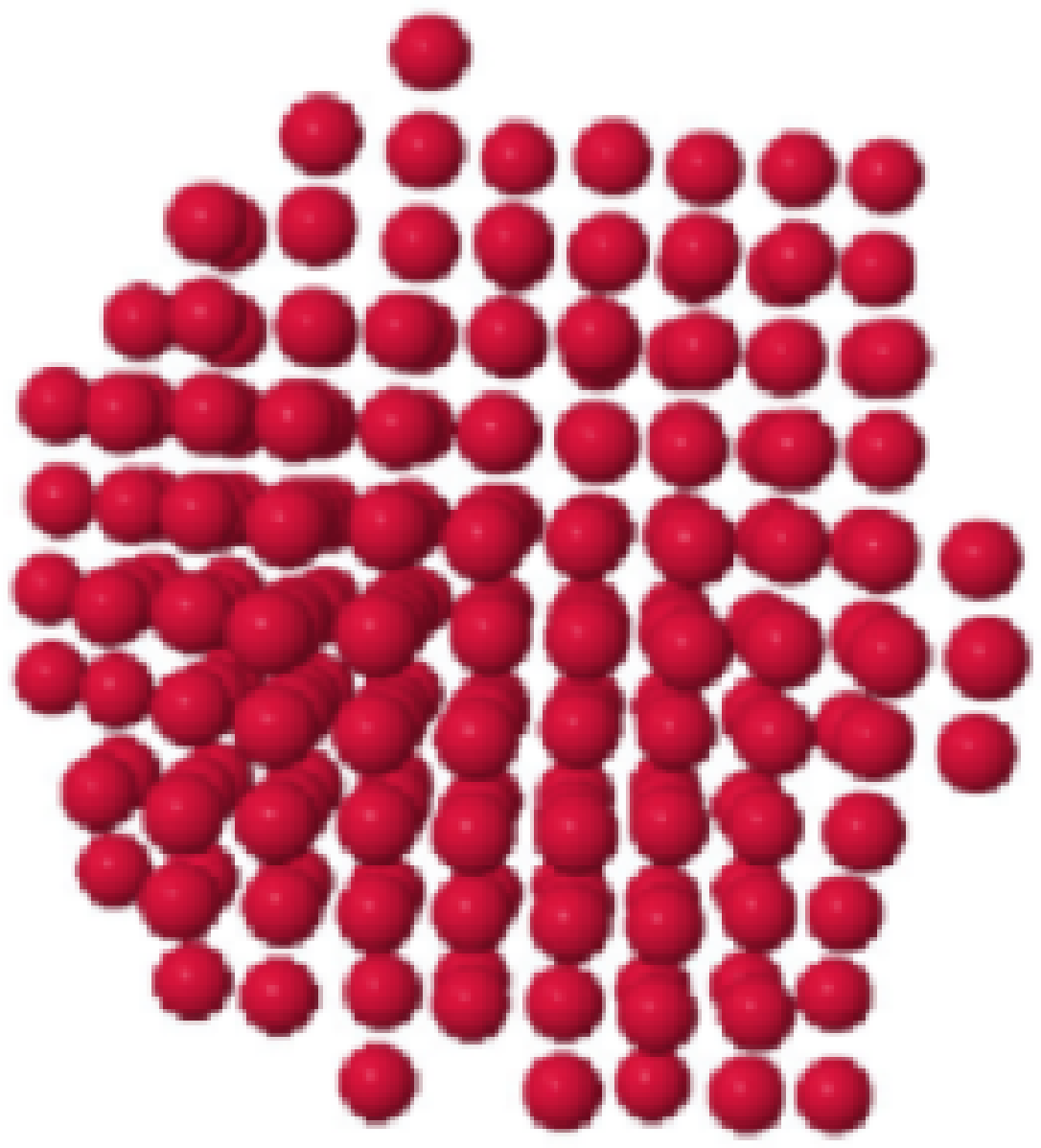} 
}
\caption{(Color online) Results of MD simulation.  216 particles interacting via
  isotropic diamond potential $V_D(r)$ self-assembles into a perfect diamond
  configuration (one configuration shown from three different
  viewpoints is shown above).  These three views clearly show that the
  result is the diamond lattice.  Closer inspection of the
  configuration shows that it is indeed tetrahedrally coordinated.}
\label{fig:config}
\end{figure}

Fifteen molecular dynamics simulations (at constant number of
particles, volume, and energy - microcanonically) were run in which
216 particles were started from a random configuration at $v=1.75694$,
at a temperature well above the freezing point (homogeneous case).
Upon slow cooling, all simulations resulted in diamond lattices with
differing numbers and types of defects, with 1 of 15 exhibiting no
defects whatsoever.  This configuration is shown in
Fig. \ref{fig:config}.  Although it may not be immediately clear from
the two-dimensional projections, this is indeed the 4-coordinated
diamond lattice, with the first neighbor at distance $a=1.045$.  The
next neighbors are at $r_2 = (\sqrt{8/3})a$ (12-coordinated),
$r_3=(\sqrt{11/3})a$ (12-coordinated), and $r_4=(\sqrt{16/3})a$
(6-coordinated), and so on.  These are the correct distances and
coordination numbers of the diamond lattice.  Heterogeneous nucleation
also produced self-assembly of the diamond lattice in a 252-particle
system, with some point defects.

In order to calculate the energy of defects (vacancies and
interstitials), molecular dynamics simulations were run in which a
particle was either removed (N=215) or inserted (N=217).  The diamond
lattice is a sub-lattice of bcc, and the interstitial was initially
set on a non-occupied bcc site (and it did not move significantly
during the relaxation).  The specific volume was set to be $v=1.75694$
in both cases.  The vacancy energy was found to be 2.866 and the
interstitial energy was 3.794.

\begin{figure}
\includegraphics[width=10cm]{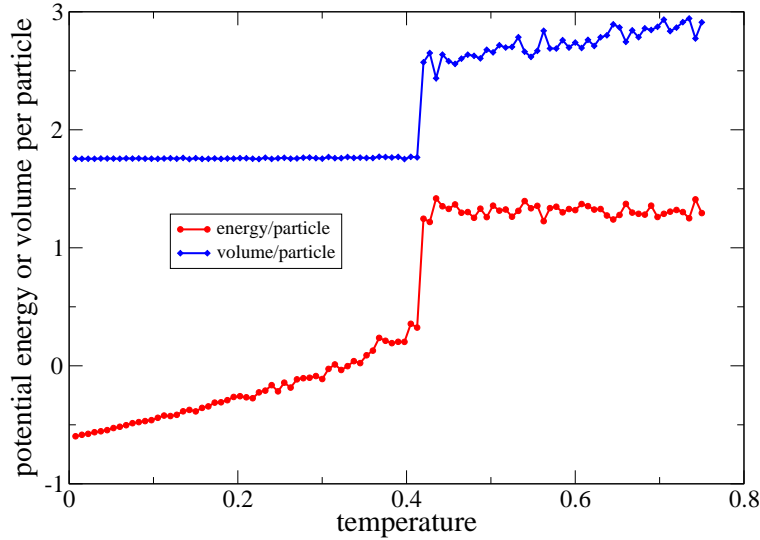}
\caption{(Color online) Volume per particle and potential energy per particle of the
melting diamond crystal, as determined from an NPT Monte Carlo simulation
(Gibbs ensemble).  The pressure here is 1.0, in units defined by
Eq. 1.  The volume curve (that above) shows that this system expands
upon melting, unlike solid water (ice), which also has a tetrahedral
structure.  Clearly this melting transition is strongly first-order.  Given in reduced units.
}
\label{fig:meltingcurve}
\end{figure}

In order to study the melting behavior of the crystal, Gibbs-ensemble
Monte Carlo simulations were run (at constant number of particles,
pressure, and temperature).  The pressure was set to be 1.0, well
within the range of stability of the lattice.  The temperature in the
simulation was slowly raised, starting near absolute zero, giving the
system sufficient time to equilibrate, and the resulting volume and
configurational energy recorded.  These two quantities, plotted in
Fig. \ref{fig:meltingcurve}, show dramatically how strongly
first-order the melting transition is.  Another interesting property
shown in this figure is that the system expands upon melting the
crystal (by roughly 45\%).  This sort of qualitative behavior is what
we would expect for a dense crystal, but perhaps not for such an open
lattice as diamond.  Indeed, solid water (ice) has a tetrahedral
structure (as do real diamond and silicon) and contracts on melting,
by approximately 8.3\% at atmospheric pressure \cite{Kauzmann}.  While
there is no intrinsic reason to believe that our model system should
act similarly, it is interesting to observe how qualitatively
differently its melting behavior is.

\begin{figure}
\includegraphics[width=10cm]{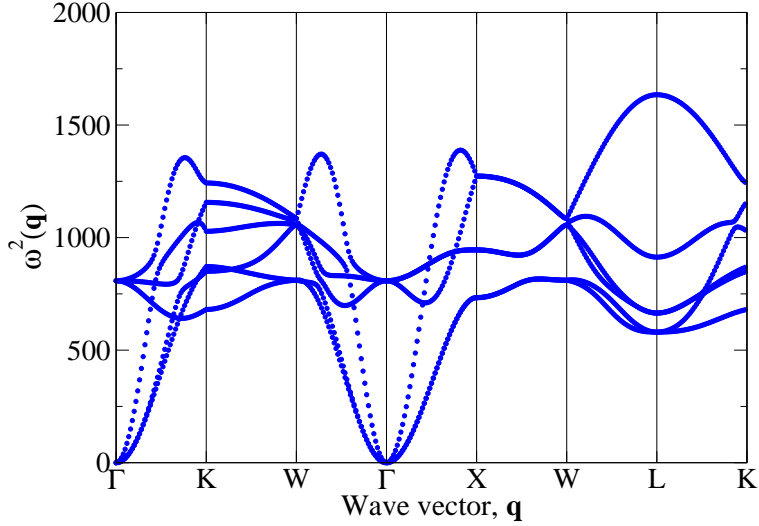}
\caption{(Color online) This is a plot of the phonon spectrum for the diamond lattice
in which particles interact via $V_D(r)$, at specific volume
$v=1.75694$.  Only certain trajectories between points of high
symmetry in the Brillouin zone are shown, although at every
wave vector, $\omega^2({\bf q}) \geq 0$.  This implies that the crystal
is linearly mechanically stable. Given in reduced units.}
\label{fig:dia-pho}
\end{figure}

The phonon spectra for the potential $V_D(r)$ with particles arranged
in a diamond lattice are shown in Fig. \ref{fig:dia-pho}.  For the
wave vectors included in the plot, all modes have real frequencies
(i.e. are propagating).  We mention that although only certain special
wave vectors are shown in the plot, at the given specific volume
$v=1.75694$, all frequencies in the three dimensional BZ are real, and
thus the crystal is linearly mechanically stable.\\

\noindent {\it B. Wurtzite}\\

\begin{figure}
  \includegraphics[width=10cm]{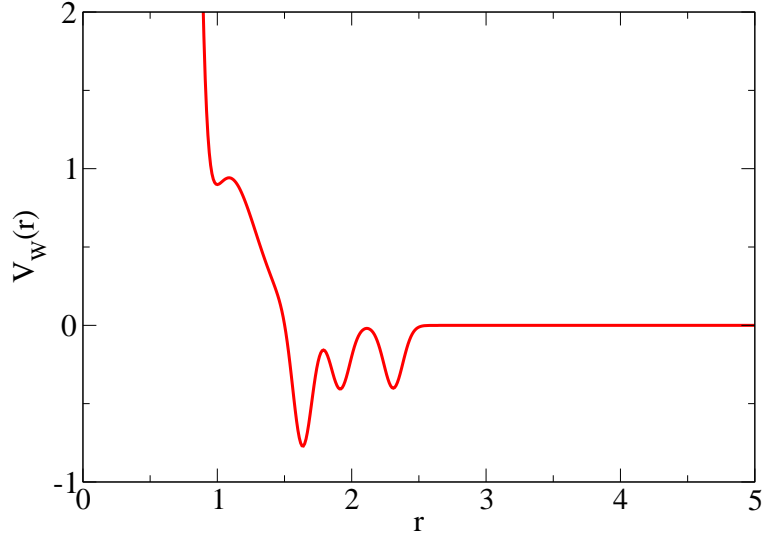}
  \caption{(Color online) Functional form of the potential, $V_W(r)$, given in
    Eq. \ref{equ:wur-pot}.  Given in reduced units.}
    \label{fig:wur-pot}
\end{figure}

The wurtzite potential, $V_W(r)$, is given by
\begin{equation}
V_W(r) =
\frac{5}{r^{12}}-\frac{6.6}{r^{10}}+4.553\exp\left[-0.6r^4\right] -
0.80\exp\left[-100(r-\sqrt{8/3})^2\right]- 
0.40\exp\left[-100(r-\sqrt{11/3})^2\right]-
0.40\exp\left[-100(r-\sqrt{16/3})^2\right].\label{equ:wur-pot}
\end{equation}
This potential is shown in Fig. \ref{fig:wur-pot}.  The
rationale for its shape is similar to that for diamond.  The most
important feature, perhaps, is the negative Gaussian centered at
$\sqrt{16/3}$, because this is comparatively close to the next
neighbor at $7/3$, which has six-fold coordination.  The analogous
neighbor in diamond is relatively separate from its next neighbor.
For wurtzite, the specific volume is $1.539601$.\\

\begin{figure}
\includegraphics[width=10cm]{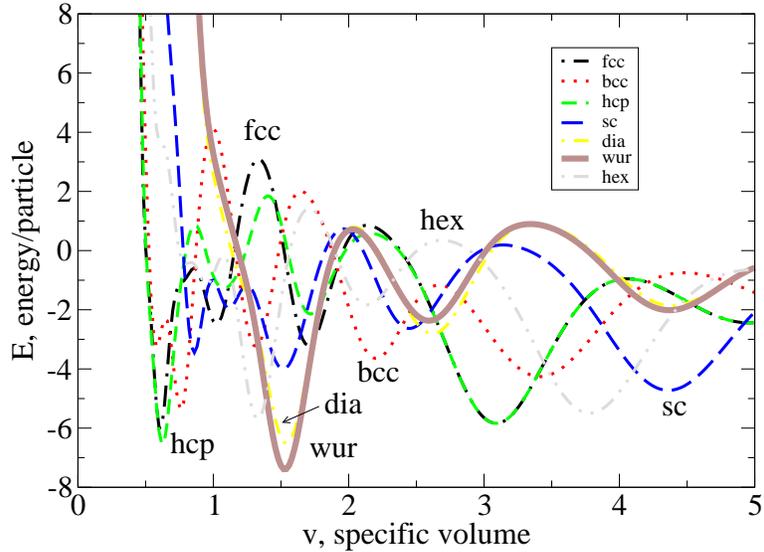}
\caption{(Color online) Lattice sums of seven lattices, including wurtzite, using the
wurtzite potential, $V_W$, given in Eq. \ref{equ:wur-pot}.  The double-tangent
construction determines that at zero temperature, the pressure range of
stability for the wurtzite lattice is 0.0 through 0.94.  Given in reduced units.  }
\label{fig:wur-lat}
\end{figure}

The lattice sums for $V_W(r)$ are given in Fig. \ref{fig:wur-lat}.
They show that there is a region of stability for the wurtzite
lattice, and using the double-tangent construction, they give the
pressure range of stability as from 0.0 through 0.94 in our
reduced units.  It is clear from the lattice sums that the HCP lattice
is an important competitor to the wurtzite (at lower specific volume).
This is natural: the wurtzite is nothing but two interlaced HCP
lattices, just as the diamond is two interlaced FCC lattices.

\begin{figure}

\subfigure[View 1: Slice of wurtzite in which its characteristic
hexagons are apparent.]  { \includegraphics[width=5cm]{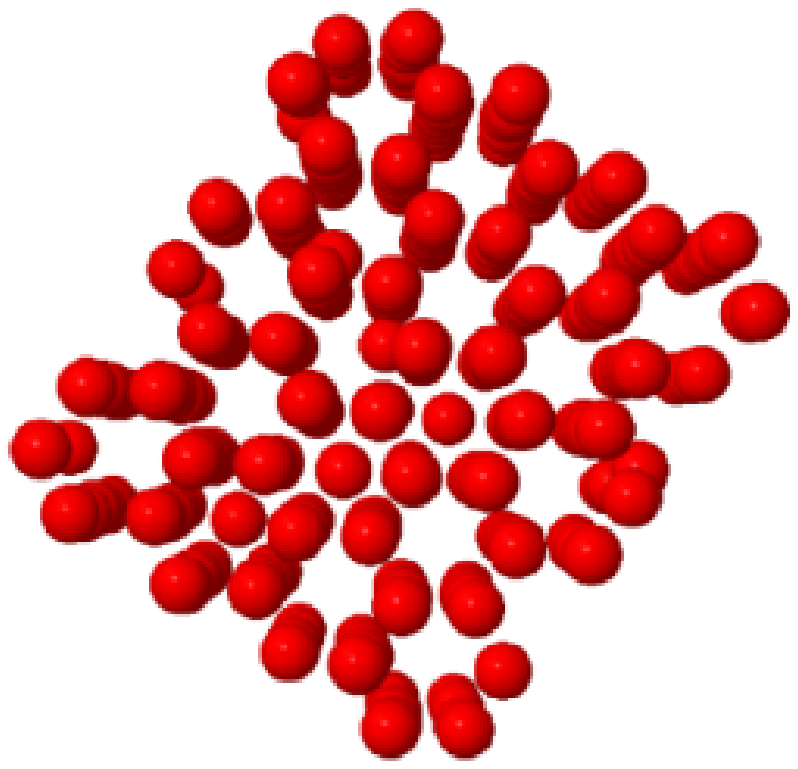}
} \subfigure[View 2: ``Zig-zags'', also characteristic of wurtzite.]
{ \includegraphics[width=5cm]{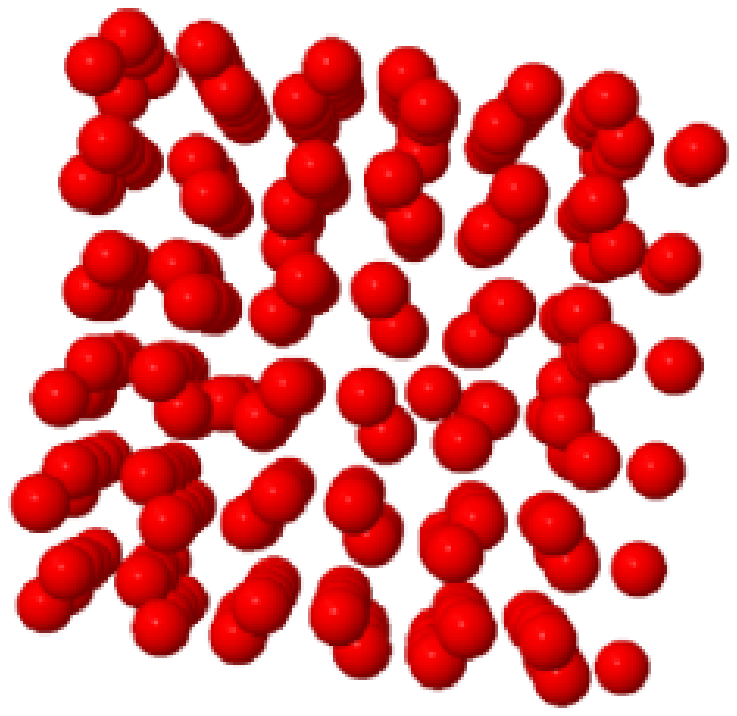} } \subfigure[View 3:
Another perspective.]  { \includegraphics[width=5cm]{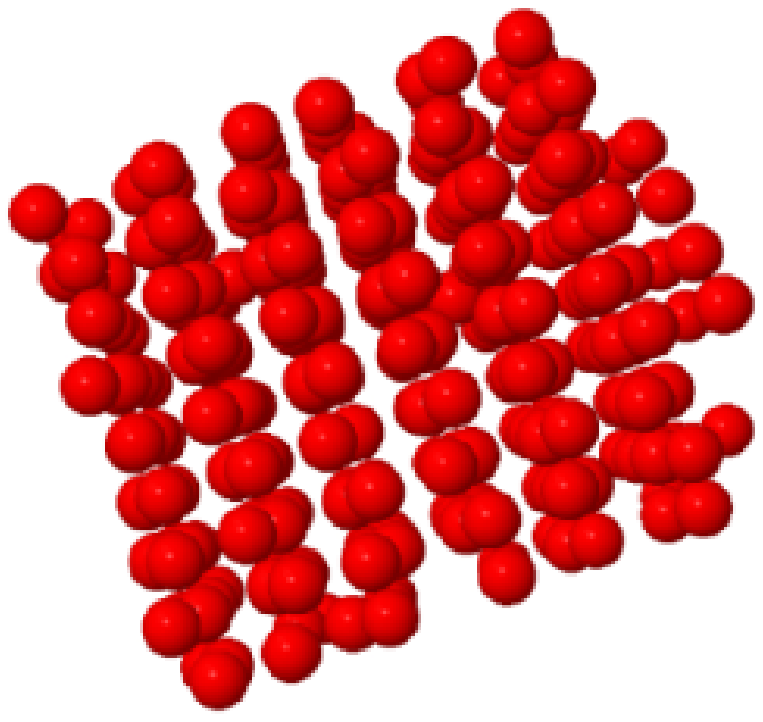}
}
\caption{(Color online) Results of MD simulation.  288 particles interacting via
  isotropic wurtzite potential $V_W(r)$ heterogeneously self-assembles
  into a near-perfect wurtzite configuration (one configuration shown
  from three different viewpoints is shown above).  These three views
  clearly show that the result is the wurtzite lattice.  Closer
  inspection of the configuration shows that it is indeed
  tetrahedrally coordinated.}
\label{fig:config-wur}
\end{figure}

The wurtzite crystal assembled upon heterogeneous, epitaxial
growth from a fixed layer of particles within a 288-particle MD
simulation, upon slow cooling.  The layer consisted of two triangular
lattice planes, one from each of the two interlaced HCP lattices, such
that growth proceeded in the c-direction (out of plane).  All other
particles in the system were inserted randomly.  The self-assembly was
not perfect, i.e. there were a number of (positive-energy) point
defects remaining in the system that could not be annealed out, but
this is to be expected for kinetic reasons.

Defect calculations were performed as in Section III.A., the diamond
case.  The vacancy energy was calculated by simply removing an atom
from the perfect wurtzite crystal and allowing the system to relax in
a microcanonical MD simulation.  We report the vacancy energy as
$2.951$, in reduced units.  For the interstitial case, consider one of
the two interlaced HCP lattices that together form wurtzite.  This is
an `ABAB...' stacking of planar triangular lattices, contrasting the
`ABCABC...' stacking of planar triangular lattices in the FCC
structure.  The wurtzite interstitial particle placed in a triangular
plane directly below the would-be ``C'' site.  Its energy is reported
as $8.889$ in reduced units.

\begin{figure}
\includegraphics[width=10cm]{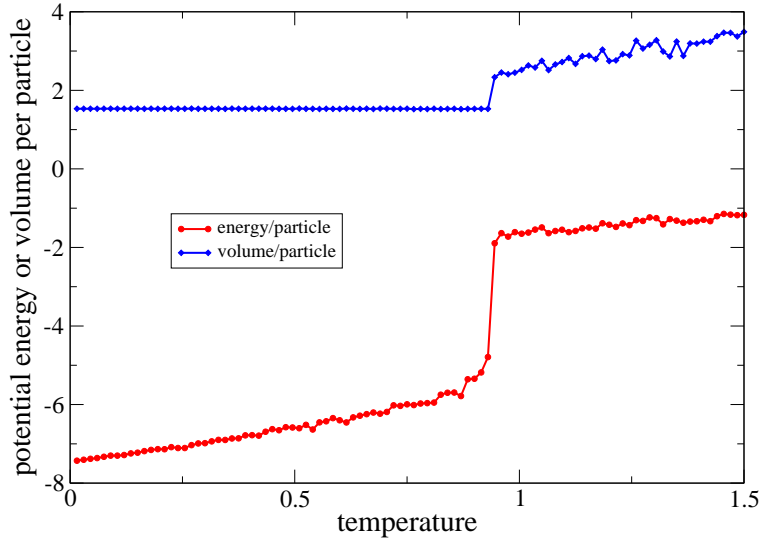}
\caption{(Color online) Volume per particle and potential energy per particle of the
melting wurtzite crystal, as determined from an NPT Monte Carlo simulation
(Gibbs ensemble).  The pressure here is 0.5, in units defined by
Eq. 1.  The volume curve (that above) shows that this system expands
upon melting, unlike solid water (ice), which also has a tetrahedral
structure.  Clearly this melting transition is strongly first-order.  Given in reduced units.
}
\label{fig:wur-meltingcurve}
\end{figure}

The melting behavior of the crystal was studied in the same way as
that of the diamond crystal in the previous section was studied,
namely using Gibbs-ensemble Monte Carlo simulations.  The pressure was
set to be 0.5, well within the range of stability of the lattice.  As
in the diamond case, the temperature in the simulation was slowly
raised, giving the system sufficient time to equilibrate, and the
resulting volume and configurational energy recorded.  These two
quantities, plotted in Fig. \ref{fig:wur-meltingcurve}, show
dramatically how strongly first-order the melting transition is for
the wurtzite case (expansion of roughly 50\%).  The melting behavior
is similar to the diamond case.  It should be noted that this is
qualitatively different from the behavior of ice-Ih, also a wurtzite
structure, which contracts upon melting.

\begin{figure}
\includegraphics[width=10cm]{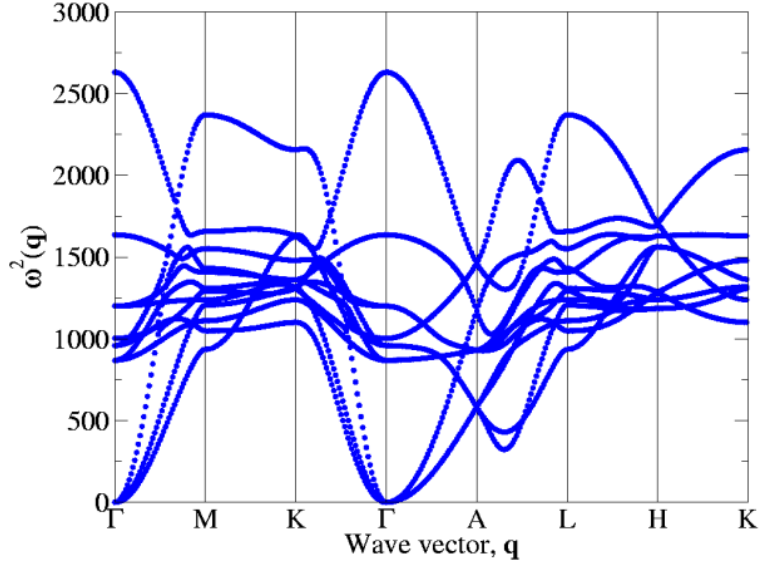}
\caption{(Color online) This is a plot of the phonon spectrum for the wurtzite
lattice in which particles interact via $V_W(r)$ at specific volume
$v=1.539601$.  Only certain trajectories between points of high
symmetry in the Brillouin zone are shown, although at every
wave vector, $\omega^2({\bf q}) \geq 0$.  This implies that the crystal
is linearly mechanically stable. Given in reduced units.}
\label{fig:wur-pho}
\end{figure}

The wurtzite phonon spectrum is shown in Fig. \ref{fig:wur-pho}.
Trajectories of high symmetry are shown in the figure, but indeed all
normal mode frequencies in the Brillouin zone are real, indicating
that the cyrstal is linearly mechanically stable.

\section{Conclusions}

In this paper, we have reported two isotropic interaction potentials
such that in an N-body classical system with particles interacting via
these potentials, the diamond and wurtzite lattices self-assemble as
the system is slowly cooled.  The functional forms of $V_D(r)$ and
$V_W(r)$ were derived using an inverse statistical mechanical
methodology developed by the authors, to be employed in tailoring
intercomponent interactions to achieve self-assembly of targeted
structures.  Admittedly, these potentials have substantial radial
complexity, but this is reminiscent of the complexity one sees in
potentials of mean force in many-body systems.  Furthermore, the
possibilities of engineering multi-feature potentials are growing
significantly as the techniques of DNA and polymer functionalization
of colloids and nanoparticles are refined.  In future work, we plan on
further exploring the self-assembly possibilities for isotropic
potentials, including other carbon structures such as graphite,
fullerenes, and nanotubes.

\begin{acknowledgements}

We would like to kindly thank Christos Likos for very helpful and
productive discussions during his visit to Princeton.  This work was
supported by the Office of Basic Energy Sciences, DOE, under Grant
No. DE-FG02-04ER46108.

\end{acknowledgements}

\end{document}